\documentclass[aps,twocolumn,prb,showpacs]{revtex4}
\usepackage{epsfig}

\begin{document}

\title{Entangled quantum currents in distant mesoscopic Josephson junctions}
\author{D. I. Tsomokos$^{1*}$, C. C. Chong$^2$, and A. Vourdas$^1$}
\affiliation{$^1$ Department of Computing, University of Bradford, Bradford,
BD7 1DP, England. \\
$^2$ Institute of High Performance Computing, 1 Science Park Road, 117528,
Singapore.}
\date{November 1, 2004; final version}

\begin{abstract}
Two mesoscopic SQUID rings which are far from each other, are considered. A
source of two-mode nonclassical microwaves irradiates the two rings with
correlated photons. The Josephson currents are in this case quantum mechanical
operators, and their expectation values with respect to the density matrix of
the microwaves, yield the experimentally observed currents. Classically
correlated (separable) and quantum mechanically correlated (entangled)
microwaves are considered, and their effect on the Josephson currents is
quantified. Results for two different examples that involve microwaves in
number states and coherent states are derived. It is shown that the quantum
statistics of the tunnelling electron pairs through the Josephson junctions in
the two rings, are correlated.
\end{abstract}

\maketitle

\section{Introduction}
Superconducting quantum interference devices (SQUIDs) exhibit quantum
coherence at the macroscopic level \cite{1}. This is a major research field
within condensed matter, and has potential applications in the developing area
of quantum information processing \cite{2,3}. A lot of the work on
superconducting rings investigates their interaction with classical
electromagnetic fields.

In the last twenty years nonclassical electromagnetic fields at low
temperatures ($k_{\rm B}T \ll \hbar\omega$) have been studied extensively
theoretically and experimentally \cite{4} at both optical and microwave
frequencies \cite{4a}. They are carefully prepared in a particular quantum
state, which is described mathematically with a density matrix $\rho$. The
interaction of SQUID rings with nonclassical microwaves has been studied in
the literature \cite{5,6}. In this case the full system, device and
microwaves, is quantum mechanical and displays interesting quantum behaviour.
For example, the quantum noise in the nonclassical microwaves affects the
Josephson currents. Experimental work, which involves the interaction of a
Josephson device with a single photon, has recently been reported
\cite{experiment}.

An important feature of two-mode nonclassical microwaves is entanglement.
Entangled electromagnetic fields have been produced experimentally \cite{6a}.
There is currently a lot of work on the classification of correlated two-mode
electromagnetic fields into classically correlated (separable) and quantum
mechanically correlated (entangled) \cite{7}. In a previous publication
\cite{8} we have studied the effects of entangled electromagnetic fields on
distant electron interference experiments. The interaction of entangled
electromagnetic fields with two superconducting charge qubits (that are
approximated by two-level systems) has recently been studied \cite{8B}. In
that work it has been shown that the entanglement is transferred from the
photons to the superconducting charge qubits. Related work has also been
reported \cite{8C}. In this paper we study the effects of entangled
electromagnetic fields on the Josephson currents of distant SQUID rings.

We consider two mesoscopic SQUID rings, which are far from each other (Fig.
1). They are irradiated with entangled microwaves, produced by a single
source. In this case the phase differences across the Josephson junctions are
quantum mechanical operators. Consequently the quantum currents, which are
sinusoidal functions of the phase differences, are also operators and their
expectation values with respect to the density matrix of the microwaves give
the observed Josephson currents. It is shown that for entangled microwaves the
currents in the two distant SQUID rings are correlated. We consider suitable
examples of separable and entangled microwaves, which differ only by
nondiagonal elements; and we show that the correlations between the induced
Josephson currents are sensitive to these nondiagonal elements.

In Sec. II we consider a single SQUID ring and study its interaction with
nonclassical microwaves. We assume the external field approximation, where the
electromagnetic field created by the Josephson current (back reaction) is
neglected. We also consider mesoscopic rings which are small in comparison to
the wavelength of the microwaves. It is shown that under these assumptions the
Josephson current is proportional to the imaginary part of the Weyl function
of the nonclassical microwaves.

In Sec. III we analyze the experiment depicted in Fig. 1, where two distant
SQUID rings are irradiated with entangled microwaves. We present examples of
separable and entangled microwaves that involve number states (Sec. IV) and
coherent states (Sec. V). In Sec. VI we present numerical calculations for
these examples. In Sec. VII we conclude with a discussion of our results.

\section{Interaction of a single SQUID ring with nonclassical microwaves}
In this section we consider a single SQUID ring and study its interaction with
both classical and nonclassical microwaves.

\subsection{Classical microwaves}
The current is $I_{\rm A}=I_1 \sin\theta_{\rm A}$, where $\theta_{\rm
A}=2e\Phi_{\rm A}$ is the phase difference across the junction due to the
total flux $\Phi_{\rm A}$ through the ring. In the external field
approximation, $\Phi_{\rm A}$ is simply the externally applied flux, while the
back reaction (i.e., the flux induced by the SQUID ring current) is neglected.
In other words the flux ${\cal L}I_{\rm A}$, where ${\cal L}$ is the
self-inductance of the ring, is assumed to be much smaller than the external
flux $\Phi_{\rm A}$. We consider a magnetic flux with a linear and a
sinusoidal component:
\begin{eqnarray}
\Phi_{\rm A}=V_{\rm A}t+\phi_{\rm A}, \;\;\; \phi_{\rm A}=A\sin(\omega_1 t).
\end{eqnarray}
In this case the current is
\begin{eqnarray}
I_{\rm A}=I_1\sin[\omega_{\rm A}t+2eA\sin(\omega_1 t)], \;\;\; \omega_{\rm
A}=2eV_{\rm A}.
\end{eqnarray}

\subsection{Nonclassical microwaves}
In this subsection we consider nonclassical electromagnetic fields, which are
carefully prepared in a particular quantum state and are described by a
density matrix $\rho$. In this case, not only the average values $\langle
E\rangle,\langle B\rangle$ of the electric and magnetic fields are known, but
also the standard deviations $\Delta E,\Delta B$ (and their higher moments).

A particular example is coherent versus squeezed microwaves. In both cases the
average values $\langle E\rangle,\langle B\rangle$ are sinusoidal functions of
time, that can be made equal for suitable values of the parameters. However
the uncertainties are $\Delta E=\Delta B= 2^{-1/2}$ for coherent microwaves;
and $\Delta E=\sigma^{-1}2^{-1/2}, \Delta B= \sigma 2^{-1/2}$, for squeezed
microwaves (where $\sigma$ is the squeezing parameter).

Another way of describing nonclassical electromagnetic fields is through the
photon counting distribution $P_N=\langle N|\rho|N\rangle$. For example, in
the case of coherent and squeezed microwaves the distribution $P_N$ is
Poissonian and sub-Poissonian, correspondingly. One of our aims in this paper
is to study how the quantum noise in the nonclassical fields, quantified by
$\Delta E,\Delta B$, or with the distribution $P_N$, affects the Josephson
currents.

In quantized electromagnetic fields the vector potential $A_i$ and the
electric field $E_i$ are dual quantum variables (operators). Strictly speaking
the dual quantum variables should be local quantities, but we consider
mesoscopic SQUID rings which are much smaller than the wavelength of the
microwaves. Therefore we can integrate these quantities over the SQUID ring
and obtain the magnetic flux and the electromotive force:
\begin{eqnarray} \label{phi_operator}
\hat\phi=\oint_C A_i dx_i, \;\;\; \hat V_{\rm EMF}=\oint_C E_i dx_i.
\end{eqnarray}

As explained above we work in the external field approximation and we neglect
the back reaction flux from the electron pairs on the external microwaves. In
this case the flux operator evolves as
\begin{eqnarray}\label{flux_operator}
\hat{\phi}(t)=\frac{\xi}{\sqrt{2}}[\hat a^{\dagger}\exp(i\omega t)+
\hat{a}\exp(-i\omega t)],
\end{eqnarray}
where $\xi$ is a parameter proportional to the area of the SQUID ring and the
$\hat{a}^{\dagger},\hat{a}$ are the photon creation and annihilation operators
(e.g., Refs. \cite{4,5}).

In order to go beyond the external field approximation we need to consider the
Hamiltonian
\begin{eqnarray}
H=\omega \left(\hat{a}^{\dagger}\hat{a} + \frac{1}{2}\right) + H_{\rm SQUID} + H_{\rm int},
\end{eqnarray}
where $H_{\rm SQUID}$ is the SQUID Hamiltonian and $H_{\rm int}$ is the
interaction term between the SQUID and the microwaves. In this case the flux
operator $\hat{\phi}'(t)$ evolves as
\begin{eqnarray}\label{flux_op_2}
\hat{\phi}'(t) &=& \exp(iHt)\hat{\phi}(0)\exp(-iHt) \nonumber \\
&=& \hat{\phi}(t)  +  \cdots
\end{eqnarray}
In this paper we work in the external field approximation and consider the
flux operator of Eq. (\ref{flux_operator}).

Consequently the phase difference $\theta_{\rm A}$ is the operator
\begin{eqnarray}\label{phase_theta_operator}
\hat{\theta}_{\rm A}=\omega_{\rm A}t + q[{\hat a}^{\dagger}\exp(i\omega t)+
\hat{a}\exp(-i\omega t)],~q=\sqrt{2}e\xi;
\end{eqnarray}
and the current also becomes an operator,
\begin{eqnarray}
\hat I_{\rm A}=I_1 \sin\{\omega_{\rm A}t + q[\hat{a}^{\dagger}\exp(i\omega t)+
\hat{a}\exp(-i\omega t)]\}.
\end{eqnarray}

Expectation values of the current are calculated by taking its trace with
respect to the density matrix $\rho$, which describes the nonclassical
electromagnetic fields,
\begin{eqnarray}\label{I_A_expression}
\langle\hat I_{\rm A}\rangle &=& \mbox{Tr}(\rho \hat I_{\rm A})=
I_1 \mbox{Im} [\exp(i\omega_{\rm A}t)\tilde W(\lambda_{\rm A})], \\
\lambda_{\rm A}&=& iq\exp(i\omega_1 t).
\end{eqnarray}
Here $\tilde W(x)$ is the Weyl function \cite{9} which is defined in terms of
the displacement operator $D(x)$ as
\begin{eqnarray}
\tilde W(x)=\mbox{Tr}[\rho D(x)] ;\;\;\; D(x)=\exp(x
\hat{a}^{\dagger}-x^{*}\hat{a}).
\end{eqnarray}
The tilde in the notation of the Weyl function indicates that it is the
two-dimensional Fourier transform of the Wigner function.

In a similar way we can calculate the $\langle {\hat I}_{\rm A}^{2}\rangle
=\mbox{Tr}(\rho {\hat I}_{\rm A}^{2})$. The second (and higher) moments of the
current describe the quantum statistics of the electron pairs tunnelling
through the Josephson junctions. As explained earlier, nonclassical
electromagnetic fields are characterized by the photon counting distribution
$P_N=\langle N|\rho|N \rangle$. The statistics of the photons threading the
ring affects the statistics of the tunnelling electron pairs, which is
quantified with the $\langle {\hat I} _{\rm A}^{2} \rangle $, $\langle {\hat
I}^{3}_{\rm A}\rangle$, etc.

\begin{figure}
\begin{center}
\scalebox{0.30}{\includegraphics{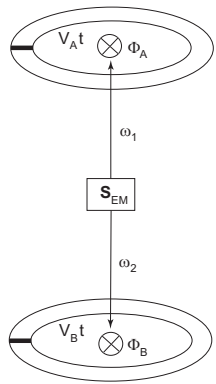}}
\end{center}
\caption{Two distant mesoscopic SQUID rings A and B are irradiated with
nonclassical microwaves of frequencies $\omega_1$ and $\omega_2$,
correspondingly. The microwaves are produced by the source ${\rm S}_{\rm EM}$
and are correlated. Classical magnetic fluxes $V_{\rm A}t$ and $V_{\rm B}t$
are also threading the two rings A and B, correspondingly.}
\end{figure}

\section{Interaction of two distant SQUID rings with entangled microwaves}
In this section we consider two SQUID rings far apart from each other, which
we refer to as A and B (Fig. 1). They are irradiated with microwaves which are
produced by the same source and are correlated. Let $\rho$ be the density
matrix of the microwaves, and
\begin{eqnarray}
\rho_{\rm A}={\rm Tr}_{\rm B}\rho,\;\;\; \rho_{\rm B}={\rm Tr}_{\rm A}\rho,
\end{eqnarray}
the density matrices of the microwaves interacting with the two SQUID rings
${\rm A}$, ${\rm B}$, correspondingly. When the density matrix $\rho$ is
factorizable as $\rho_{\rm fact}=\rho_{\rm A}\otimes\rho_{\rm B}$ the two
modes are not correlated. If it can be written as  $\rho_{\rm sep}=\sum_i p_i
\rho_{{\rm A}i}\otimes \rho_{{\rm B}i}$, where $p_i$ are probabilities, it is
called separable and the two modes are classically correlated. Density
matrices which cannot be written in one of these two forms are entangled
(quantum mechanically correlated). There has been a lot of work on criteria
which distinguish separable and entangled states \cite{7}.

The currents in the two SQUIDs are
\begin{eqnarray} \langle \hat I_{\rm
A}\rangle &=& I_{1} \mbox{Tr}(\rho_{\rm A}\sin \hat
\theta_{\rm A}), \label{I_A} \\
\langle \hat I_{\rm B}\rangle &=& I_{2}
\mbox{Tr}(\rho_{\rm B}\sin \hat \theta_{\rm B}). \label{I_B}
\end{eqnarray}
The $\langle \hat I_{\rm A}\rangle$ is written in terms of the Weyl function
$\tilde W(\lambda_{\rm A})$ in Eq. (\ref{I_A_expression}), and similarly for B
one may obtain $\langle \hat I_{\rm B}\rangle = I_2 \mbox{Im}
[\exp(i\omega_{\rm B}t)\tilde W(\lambda_{\rm B})]$, where $\lambda_{\rm
B}=iq\exp(i\omega_2t)$ and $\omega_{\rm B}=2eV_{\rm B}$.

The expectation value of the product of the two current operators is given by:
\begin{equation}\label{I_AB}
\langle\hat I_{\rm A}\hat I_{\rm B}\rangle = I_{1}I_{2} \mbox{Tr}(\rho
\sin\hat\theta_{\rm A}\sin\hat\theta_{\rm B}).
\end{equation}
We consider the ratio of the currents
\begin{equation}\label{ratio}
R = \frac{\langle\hat I_{\rm A} \hat I_{\rm B}\rangle} {\langle\hat I_{\rm A}\rangle
\langle\hat I_{\rm B}\rangle}.
\end{equation}
For factorizable density matrices $\rho_{\rm fact}=\rho_{A}\otimes\rho_B$ we
easily see that $R_{\rm fact}=1$. For separable density matrices $\rho_{\rm
sep}=\sum_i p_i \rho_{{\rm A}i}\otimes \rho_{{\rm B}i}$ we get
\begin{eqnarray}
R_{\rm sep} = \frac{\sum_i p_i \langle \hat I_{{\rm A}i}\rangle\langle \hat
I_{{\rm B}i}\rangle} {(\sum_k p_k \langle \hat I_{{\rm A}k}\rangle)(\sum_l p_l
\langle \hat I_{{\rm B}l}\rangle)}.
\end{eqnarray}
We also calculate the second moments
\begin{eqnarray}
\langle {\hat I}_{\rm A}^2 \rangle &=& I_{1}^2 \mbox{Tr}[\rho_{\rm A}(\sin
\hat\theta_{\rm A})^2], \label{I_A_2} \\
\langle {\hat I}_{\rm B}^2\rangle &=& I_{2}^2 \mbox{Tr}[\rho_{\rm B}(\sin \hat
\theta_{\rm B})^2]. \label{I_B_2}
\end{eqnarray}
As explained earlier, the statistics of the photons threading the ring affects
the statistics of the tunnelling electron pairs, which is quantified with the
$\langle {\hat I}_{\rm A} {\hat I}_{\rm B}\rangle$, $\langle {\hat I} _{\rm
A}^{2} \rangle $, $\langle {\hat I}^{2}_{\rm B}\rangle$, etc.

In the following sections we consider particular examples for the density
matrix $\rho$ of the nonclassical microwaves that interact with the two SQUID
rings, and examine its effect on these quantities.

\section{Microwaves in number states}
We consider microwaves in the separable (mixed) state
\begin{equation}\label{num_1_sep}
\rho_{\rm sep}=\frac{1}{2}(|N_1 N_2 \rangle \langle N_1 N_2| + |N_2N_1\rangle
\langle N_2N_1|),
\end{equation}
where $N_1\neq N_2$. We also consider microwaves in the entangled state
$|s\rangle=2^{-1/2}(|N_1N_2\rangle+|N_2N_1\rangle)$, which is a pure state.
The density matrix of $|s\rangle$ is
\begin{equation}\label{num_1_ent}
\rho_{\rm ent}=\rho_{\rm sep} + \frac{1}{2}(|N_1 N_2 \rangle \langle N_2
N_1|+|N_2 N_1 \rangle \langle N_1 N_2|),
\end{equation}
where the $\rho_{\rm sep}$ is given by Eq. (\ref{num_1_sep}). It is seen that
the $\rho_{\rm ent}$ and the $\rho_{\rm sep}$ differ only by the above
nondiagonal elements, and below we calculate their effect on the Josephson
currents.

We note that it is possible to have `interpolating' density matrices of the
form
\begin{eqnarray}\label{interp}
\rho _{\rm p} &=& p\rho_{\rm sep}+(1-p)\rho_{\rm ent} \\
 &=& \rho_{\rm sep} + \frac{1-p}{2}(|N_1 N_2 \rangle \langle N_2 N_1|+|N_2 N_1
\rangle \langle N_1 N_2|) \nonumber
\end{eqnarray}
where $0\le p\le 1$. Below we present results for the two extreme cases of
$\rho_{\rm sep}$, where the nondiagonal terms make no contribution; and for
the $\rho_{\rm ent}$, where the nondiagonal terms make maximal contribution.
We also present numerical results for the case of $\rho_{\rm p}$.

In this example, the reduced density matrices are the same for both the
separable and entangled states:
\begin{eqnarray}\label{reduced_rho}
\rho_{\rm sep, A}&=&\rho_{\rm ent, A}=\rho_{\rm sep, B}=\rho_{\rm ent, B} \\ \nonumber
&=& \frac{1}{2} (|N_1\rangle\langle N_1| + |N_2\rangle\langle N_2|).
\end{eqnarray}
Consequently in this example $\langle\hat I_{\rm A}\rangle_{\rm
sep}=\langle\hat I_{\rm A}\rangle_{\rm ent}$, and also $\langle\hat I_{\rm
B}\rangle_{\rm sep}=\langle\hat I_{\rm B}\rangle_{\rm ent}$.

\subsection{Classically correlated photons}
For the density matrix $\rho_{\rm sep}$ of Eq. (\ref{num_1_sep}) we find
\begin{eqnarray}
\langle\hat I_{\rm A}\rangle &=& I_{1} C \sin(\omega_{\rm A}t); \label{I_reduced_A_num}\\
\langle\hat I_{\rm B}\rangle &=& I_{2} C \sin(\omega_{\rm B}t); \label{I_reduced_B_num}\\
C&=&\frac{1}{2}\exp\left(-\frac{q^2}{2}\right)[L_{N_1}(q^2)+L_{N_2}(q^2)],\label{C}
\end{eqnarray}
where the $L_{n}^{\alpha}(x)$ are Laguerre polynomials (in the case of Eq.
(\ref{C}) we have $\alpha=0$). The currents $\langle\hat I_{\rm A}\rangle,
\langle\hat I_{\rm B}\rangle$ are in this case independent of the microwave
frequencies $\omega_1,\omega_2$.

The expectation value of the product of the two currents [Eq. (\ref{I_AB})] is
\begin{eqnarray}
\langle{\hat I}_{\rm A} {\hat I}_{\rm B}\rangle_{\rm sep}&=& I_{1}I_{2} C_1
\sin(\omega_{\rm A}t)\sin(\omega_{\rm B}t), \label{I_sep} \\
C_1 &=& \exp(-q^2) L_{N_1}(q^2)L_{N_2}(q^2). \label{C_1}
\end{eqnarray}
Consequently the ratio $R$ of Eq. (\ref{ratio}) is
\begin{equation}\label{R_sep}
R_{\rm sep}=\frac{C_1}{C^2}= \frac{4L_{N_1}(q^2)L_{N_2}(q^2)}
{[L_{N_1}(q^2)+L_{N_2}(q^2)]^2}.
\end{equation}
In this example the $R_{\rm sep}$ is time-independent.

The moments of the currents, defined by Eqs. (\ref{I_A_2}),
(\ref{I_B_2}), are also calculated:
\begin{eqnarray}
\langle \hat I_{\rm A}^2 \rangle &=&  \frac{I_{1}^2}{2}
[1 - C_2 \cos(2\omega_{\rm A}t) ], \label{I_A_squared} \\
\langle \hat I_{\rm B}^2 \rangle &=&  \frac{I_{2}^2}{2}
[1 - C_2 \cos(2\omega_{\rm B}t)], \label{I_B_squared} \\
C_2 &=& \frac{1}{2}\exp(-2q^2)[L_{N_1}(4q^2)+L_{N_2}(4q^2)].
\end{eqnarray}

\subsection{Quantum mechanically correlated photons}
For the case of $\rho_{\rm ent}$ the reduced density matrices $\rho_{\rm
A},\rho_{\rm B}$ are those given by Eq. (\ref{reduced_rho}), and consequently
the $\langle\hat I_{\rm A}\rangle,\langle\hat I_{\rm B}\rangle$ are the same
as in Eqs. (\ref{I_reduced_A_num}), (\ref{I_reduced_B_num}); and the
$\langle\hat I_{\rm A}^2\rangle,\langle\hat I_{\rm B}^2\rangle$ are the same
as in Eqs. (\ref{I_A_squared}), (\ref{I_B_squared}).

However in this case the $\langle{\hat I}_{\rm A}{\hat I}_{\rm B}\rangle$ is
\begin{equation}\label{I_ent}
\langle{\hat I}_{\rm A}{\hat I}_{\rm B}\rangle_{\rm ent} = \langle{\hat
I}_{\rm A}{\hat I}_{\rm B}\rangle_{\rm sep} + I_{\rm cross},
\end{equation}
where
\begin{eqnarray}
\lefteqn{I_{\rm cross} = -I_{1}I_{2}C_3
\cos[(N_1-N_2)(\omega_1-\omega_2)t] } \nonumber \\
&& \times  [\cos(\omega_{\rm A}t+\omega_{\rm B}t)-(-1)^{N_1-N_2}
\cos(\omega_{\rm A}t-\omega_{\rm B}t)],
\nonumber \\ \label{I_cross}\\
&&C_3=\frac{1}{2} \exp(-q^2) L_{N_1}^{N_2-N_1}(q^2)L_{N_2}^{N_1-N_2}(q^2).
\label{C_3}
\end{eqnarray}
The term $I_{\rm cross}$ is induced by the nondiagonal elements of $\rho_{\rm
ent}$, and depends on the photon frequencies $\omega_1,\omega_2$. This term
quantifies the difference between the effect of separable and entangled
microwaves on the Josephson currents. We note that the nondiagonal terms of
$\rho_{\rm ent}$ [Eq. (\ref{num_1_ent})] are small and consequently they are
very sensitive to the back reaction. Therefore our results which neglect the
back reaction are relevant to experiments with small Josephson currents. In
other words it is required that the fluxes ${\cal L}_{\rm A}I_{\rm A}$ and
${\cal L}_{\rm B}I_{\rm B}$ are much smaller than the external flux.

The ratio $R$ of Eq. (\ref{ratio}) can be simplified in two distinct
expressions according to whether the difference $N_1-N_2$ is even or odd. In
the case $N_1-N_2=2k$, the ratio is
\begin{equation}\label{R_ent_even}
R_{\rm ent}^{(2k)}=R_{\rm sep}+\frac{4L_{N_1}^{-2k}(q^2)L_{N_2}^{2k}(q^2)}
{[L_{N_1}(q^2)+L_{N_2}(q^2)]^2}\cos(\Omega t),
\end{equation}
where
\begin{equation}\label{Omega}
\Omega=(N_1-N_2)(\omega_1-\omega_2).
\end{equation}
It is seen that the $R_{\rm ent}^{(2k)}$ oscillates around the $R_{\rm sep}$
with frequency $\Omega$ given by Eq. (\ref{Omega}). If there is no detuning
between the nonclassical electromagnetic fields, i.e. $\omega_1=\omega_2$,
then $R_{\rm ent}^{(2k)}$ is constant, although it is still $R_{\rm ent}\neq
R_{\rm sep}$.

In the case $N_1-N_2=2k+1$ the ratio is
\begin{eqnarray}\label{R_ent_odd}
\lefteqn{R_{\rm ent}^{(2k+1)}= R_{\rm sep} } \nonumber \\
&& -\frac{4L_{N_1}^{-2k-1}(q^2)L_{N_2}^{2k+1}(q^2)}
{[L_{N_1}(q^2)+L_{N_2}(q^2)]^2}\frac{\cos(\Omega t)}{\tan(\omega_{\rm
A}t)\tan(\omega_{\rm B}t)}.
\end{eqnarray}
In both cases the $R_{\rm ent}$ is time-dependent and it is a function of the
photon frequencies $\omega_1,\omega_2$, in contrast to the case of $R_{\rm
sep}$ (which is time-independent).

\section{Microwaves in coherent states}
We consider microwaves in the classically correlated state
\begin{eqnarray}\label{rho_sep_coherent}
\rho_{{\rm sep}}=\frac{1}{2}(|A_1 A_2\rangle \langle A_1 A_2| +|A_2 A_1\rangle
\langle A_2 A_1|).
\end{eqnarray}
The $|A_1\rangle$, $|A_2\rangle$ are microwave coherent states (eigenstates of
the annihilation operators). We also consider the entangled state $|u
\rangle={\cal N}(|A_1 A_2\rangle +|A_2 A_1\rangle)$, with density matrix
\begin{eqnarray}\label{rho_ent_coherent}
\rho_{\rm ent}&=& 2{\cal N} ^2 \rho_{\rm sep} \nonumber \\
&+& {\cal N}^2 (|A_1 A_2\rangle \langle A_2 A_1| +|A_2 A_1\rangle \langle A_1 A_2|),
\end{eqnarray}
where the normalization constant is given by
\begin{eqnarray}\label{normalization_N}
{\cal N}=\left[2+2\exp\left(-|A_1-A_2|^2\right)\right]^{-1/2}.
\end{eqnarray}

\subsection{Classically correlated photons}
For microwaves in the separable state of Eq. (\ref{rho_sep_coherent}) the
reduced density matrices are
\begin{eqnarray}\label{reduced_rho_sep}
\rho_{\rm sep,A}=\rho_{\rm sep,B}=\frac{1}{2}(|A_1\rangle \langle A_1|
+|A_2\rangle\langle A_2|),
\end{eqnarray}
and hence the currents in A and B are
\begin{eqnarray} \label{I_A_sep}
\langle\hat I_{\rm A}\rangle_{\rm sep} &=& \frac{I_{1}}{2}\exp(-\frac{q^2}{2})
\{\sin[\omega_{\rm A}t+2q|A_1|\cos(\omega_1 t-\theta_{1})] \nonumber \\
&+& \sin[\omega_{\rm A}t+2q|A_2|\cos(\omega_1 t-\theta_{2})]\},
\end{eqnarray}
\begin{eqnarray} \label{I_B_sep}
\langle\hat I_{\rm B}\rangle_{\rm sep} &=& \frac{I_{2}}{2}\exp(-\frac{q^2}{2})
\{\sin[\omega_{\rm B}t+2q|A_1|\cos(\omega_2 t-\theta_{1})] \nonumber \\
&+& \sin[\omega_{\rm B}t+2q|A_2|\cos(\omega_2 t-\theta_{2})]\},
\end{eqnarray}
where $\theta_1=\arg (A_1)$, and $\theta_2=\arg (A_2)$. We have also
calculated numerically the ratio $R_{\rm sep}$.

\subsection{Quantum mechanically correlated photons}
For microwaves in the entangled state of Eq. (\ref{rho_ent_coherent}) the
reduced density matrices are
\begin{eqnarray}\label{reduced_rho_ent}
\rho_{\rm ent,A}&=&\rho_{\rm ent,B}= {\cal N}^2(|A_1\rangle \langle A_1|
+|A_2\rangle \langle A_2| \nonumber \\
&+& \tau |A_1\rangle\langle A_2| + \tau^{*} |A_2\rangle\langle A_1|),
\end{eqnarray}
where
\begin{eqnarray}
\tau = \langle A_1|A_2\rangle = \exp\left(-\frac{|A_1|^2}{2}
-\frac{|A_2|^2}{2} + A_1^{*} A_2 \right).
\end{eqnarray}

The current in A is
\begin{eqnarray}
\langle\hat I_{\rm A}\rangle_{\rm ent}= 2{\cal N}^2 \langle\hat I_{\rm
A}\rangle_{\rm sep}+ {\cal N}^2 E F_1 \exp\left(-\frac{q^2}{2}\right)I_{1},
\end{eqnarray}
where
\begin{equation}
E = \exp[-|A_1|^2 - |A_2|^2+2|A_1 A_2|\cos(\theta_{1}-\theta_{2})],
\end{equation}
\begin{eqnarray}
\lefteqn{ F_1 =\{ \exp[q|A_1|S_{A,1}(t)-q|A_2|S_{A,2}(t)]} \nonumber \\
&& + \exp[-q|A_1|S_{A,1}(t)+q|A_2|S_{A,2}(t)]\} \nonumber \\
&& \times\sin[\omega_A t+q|A_1|C_{A,1}(t)+q|A_2|C_{A,2}(t)], \label{I_A_ent}
\end{eqnarray}
and
\begin{eqnarray}
S_{A,1}&=&\sin(\omega_1 t-\theta_{1}), \;\;\;
S_{A,2}=\sin(\omega_1 t-\theta_{2}), \nonumber \\
C_{A,1}&=&\cos(\omega_1 t-\theta_{1}), \;\;\;
C_{A,2}=\cos(\omega_1t-\theta_{2}).
\end{eqnarray}
The current in B is
\begin{eqnarray}
\langle\hat I_{\rm B}\rangle_{\rm ent} =2{\cal N}^2 \langle\hat I_{\rm
B}\rangle_{\rm sep}+ {\cal N}^2 E F_2 \exp\left(-\frac{q^2}{2}\right)I_{2},
\end{eqnarray}
where
\begin{eqnarray}
\lefteqn{ F_2=\{\exp[q|A_1|S_{B,1}(t)-q|A_2|S_{B,2}(t)]} \nonumber \\
&& + \exp[-q|A_1|S_{B,1}(t)+q|A_2|S_{B,2}(t)] \} \nonumber \\
&& \times\sin[\omega_B t+q|A_1|C_{B,1}(t)+q|A_2|C_{B,2}(t)],\label{I_B_ent}
\end{eqnarray}
and
\begin{eqnarray}
S_{B,1}&=&\sin(\omega_2 t-\theta_{1}), \;\;\; S_{B,2}=\sin(\omega_2 t-\theta_{2}),\nonumber \\
C_{B,1}&=&\cos(\omega_2t-\theta_{1}), \;\;\;
C_{B,2}=\cos(\omega_2t-\theta_{2}).
\end{eqnarray}
We have also calculated numerically the ratio $R_{\rm ent}$.

As we already explained, the nondiagonal terms in $\rho_{\rm ent}$ are very
sensitive to back reaction and therefore these results are relevant to
experiments with small Josephson currents. In other words it is required that
the fluxes ${\cal L}_{\rm A}I_{\rm A}$ and ${\cal L}_{\rm B}I_{\rm B}$ are
much smaller than the external flux.

\section{Numerical results}
In all numerical results of Figs. 2 to 6 the microwave frequencies are
$\omega_1=1.2\times 10^{-4},\omega_2=10^{-4}$, in units where $k_B=\hbar=c=1$.
The critical currents are $I_{1}=I_{2}=1$. The other parameters are $\xi=1$,
$\omega_{\rm A}=\omega_1$, $\omega_{\rm B}=\omega_2$, $N_1=2$, $N_2=0$, and
the arguments of the coherent eigenstates are $\theta_1=\theta_2=0$. For a
meaningful comparison between microwaves in number states and microwaves in
coherent states, we take $|A_1|^2=N_1$ and $|A_2|^2=N_2$, so that the average
number of photons in the coherent states is equal to the number of photons in
the number states.

In Fig. 2 we plot $R_{\rm sep}$ against $(\omega_1-\omega_2)t$ for currents
induced by microwaves in the number state of Eq. (\ref{num_1_sep}) with
$N_1=2,N_2=0$ (line of circles), and the coherent state of Eq.
(\ref{rho_sep_coherent}) with $A_1=\sqrt{2},A_2=0$ (solid line). It is seen
that two different microwave states with the same average number of photons
give different results on the quantum statistics of the electron pairs.

In Fig. 3 we plot $R_{\rm sep}-R_{\rm ent}$ against $(\omega_1-\omega_2)t$ for
currents induced by microwaves in (a) the number states of Eqs.
(\ref{num_1_sep}) and (\ref{num_1_ent}) with $N_1=2,N_2=0$, and (b) the
coherent states of Eqs. (\ref{rho_sep_coherent}) and (\ref{rho_ent_coherent})
with $A_1=\sqrt{2},A_2=0$. It is seen that the separable and entangled states,
which differ only by nondiagonal elements, give different results. As
expected, the difference (which shows the effect of the nondiagonal elements)
is small, but it is nonzero.

In Fig. 4 we plot (a) $\langle {\hat I}_{\rm A}\rangle_{\rm sep} - \langle
{\hat I}_{\rm A} \rangle_{\rm ent}$, and (b) $\langle {\hat I}_{\rm A}^2
\rangle_{\rm sep} - \langle {\hat I}_{\rm A}^2 \rangle_{\rm ent}$, against
$(\omega_1-\omega_2)t$ for microwaves in the coherent state $\rho_{\rm A,sep}$
of Eq. (\ref{reduced_rho_sep}) and $\rho_{\rm A,ent}$ of Eq.
(\ref{reduced_rho_ent}) with $A_1=\sqrt{2},A_2=0$. For coherent states
$\rho_{\rm ent,A}$ is not equal to $\rho_{\rm sep,A}$ [cf. Eqs.
(\ref{reduced_rho_ent}), (\ref{reduced_rho_sep})] and consequently the
corresponding currents are different. For number states, the currents
corresponding to $\rho_{\rm ent,A}$ and $\rho_{\rm sep,A}$ are the same
because $\rho_{\rm ent,A}=\rho_{\rm sep,A}$ [cf. Eq. (\ref{reduced_rho})].

In Fig. 5 we plot $\langle {\hat I}_{\rm A} {\hat I}_{\rm B}\rangle_{\rm sep}
- \langle {\hat I}_{\rm A} {\hat I}_{\rm B} \rangle_{\rm ent}$ for (a) number
states of Eqs. (\ref{num_1_sep}) and (\ref{num_1_ent}) with $N_1=2,N_2=0$, and
(b) coherent states of Eqs. (\ref{rho_sep_coherent}) and
(\ref{rho_ent_coherent}) with $A_1=\sqrt{2},A_2=0$, against
$(\omega_1-\omega_2)t$. In this figure also, we get different results due to
the nondiagonal elements in the entangled state.

In Fig. 6 we plot the ratio $R_{\rm p}$ of Eq. (\ref{ratio}) against $p\in
[0,1]$ for currents that are induced by the interpolating density matrix
$\rho_{\rm p}$ of Eq. (\ref{interp}) for number states with $N_1=2,N_2=0$. The
time $t$ has been fixed so that $(\omega_1-\omega_2)t=\pi/2$.

\section{Discussion}
We have considered the interaction of SQUID rings with nonclassical
microwaves. We have assumed the external field approximation, where the
electromagnetic field created by the Josephson currents (back reaction) is
neglected. We have also considered small rings in comparison to the wavelength
of the microwaves and taken as dual quantum variables the magnetic flux and
electromotive force of Eq. (\ref{phi_operator}).

The Josephson current is an operator and its expectation value with respect to
the density matrix of the nonclassical microwaves is the observed current. We
have shown that the expectation value of the current is proportional to the
imaginary part of the Weyl function [Eq. (\ref{I_A_expression})]. This shows
clearly how the full density matrix of the microwaves affects the Josephson
current. The higher moments of the current $\langle {\hat I}_{\rm A}^{M}
\rangle$ can also be calculated and used to quantify the statistics of the
tunnelling electron pairs. It has been shown that the statistics of the
irradiating photons determine the tunnelling statistics of the electron pairs.

We have also considered the interaction of two distant SQUID rings A and B
with two-mode nonclassical microwaves, which are produced by the same source.
It has been shown that classically correlated (separable) and quantum
mechanically correlated (entangled) photons induce different Josephson
currents and different tunnelling statistics in the two devices. The results
show that the entangled photons produce entangled Josephson currents in the
distant SQUID rings. This can have applications in the general area of quantum
information processing.

The work can be extended in various directions. The first is to take into
account the back reaction and include the extra terms which we have neglected
in Eq. (\ref{flux_op_2}). This can be done numerically. The second direction
is the study of Bell-like inequalities for the Josephson currents; which are
violated when the currents are entangled. Another direction is the potential
use of the system as a detector of entangled photons. There is a lot of work
on the use of mesoscopic devices as detectors \cite{detector}. Application of
the present work in this direction requires further work, which could lead to
the development of a detection system for entangled photons based on two
distant SQUID rings.

$\\$
-----------------------------
$\\$ $^*$ \textbf{Corresponding author:} d.i.tsomokos@physics.org

\newpage

\begin{figure}
\begin{center}
\scalebox{0.4}{\includegraphics{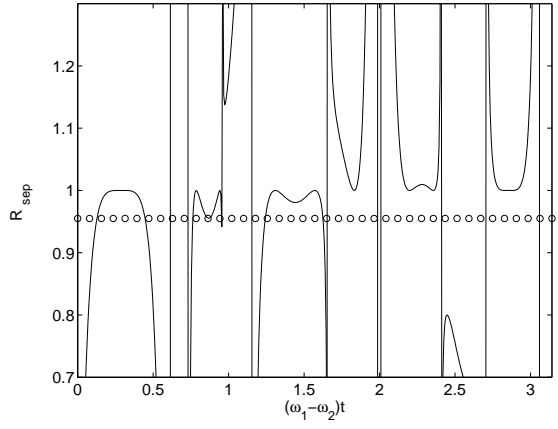}}
\end{center}
\caption{$R_{\rm sep}$ against $(\omega_1-\omega_2)t$ for the number state of
Eq. (\ref{num_1_sep}) with $N_1=2,N_2=0$ (line of circles), and the coherent
state of Eq. (\ref{rho_sep_coherent}) with $A_1=\sqrt{2},A_2=0$ (solid line).
The photon frequencies are $\omega_1=1.2\times 10^{-4}$ and
$\omega_2=10^{-4}$, in units where $k_B=\hbar=c=1$.}
\end{figure}
\begin{figure}
\begin{center}
\scalebox{0.4}{\includegraphics{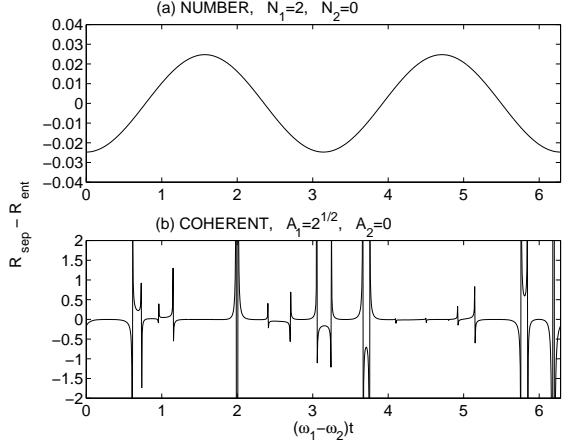}}
\end{center}
\caption{$R_{\rm sep}-R_{\rm ent}$ against $(\omega_1-\omega_2)t$ for (a)
number states of Eqs. (\ref{num_1_sep}) and (\ref{num_1_ent}) with
$N_1=2,N_2=0$, and (b) coherent states of Eqs. (\ref{rho_sep_coherent}) and
(\ref{rho_ent_coherent}) with $A_1=\sqrt{2},A_2=0$. The photon frequencies are
$\omega_1=1.2\times 10^{-4}$ and $\omega_2=10^{-4}$, in units where
$k_B=\hbar=c=1$.}
\end{figure}
\begin{figure}
\begin{center}
\scalebox{0.4}{\includegraphics{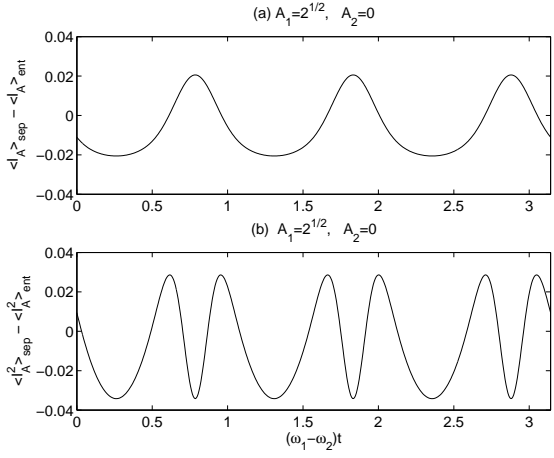}}
\end{center}
\caption{(a) $\langle {\hat I}_{\rm A}\rangle_{\rm sep} - \langle {\hat
I}_{\rm A} \rangle_{\rm ent}$, and (b) $\langle {\hat I}_{\rm A}^2
\rangle_{\rm sep} - \langle {\hat I}_{\rm A}^2 \rangle_{\rm ent}$ against
$(\omega_1-\omega_2)t$ for the coherent state $\rho_{\rm sep,A}$ of Eq.
(\ref{reduced_rho_sep}) and $\rho_{\rm ent,A}$ of Eq. (\ref{reduced_rho_ent})
with $A_1=\sqrt{2},A_2=0$. The photon frequencies are $\omega_1=1.2\times
10^{-4}$ and $\omega_2=10^{-4}$, in units where $k_B=\hbar=c=1$.}
\end{figure}
\begin{figure}
\begin{center}
\scalebox{0.4}{\includegraphics{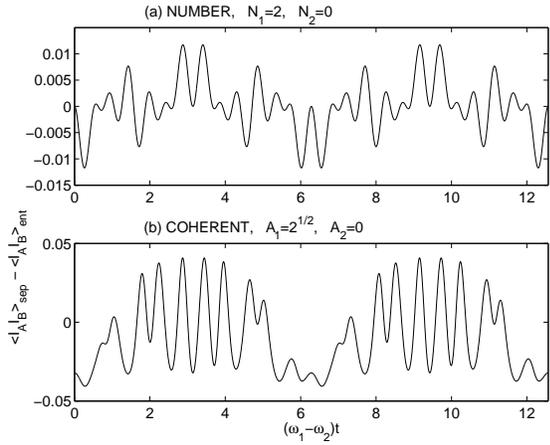}}
\end{center}
\caption{$\langle {\hat I}_{\rm A} {\hat I}_{\rm B}\rangle_{\rm sep} - \langle
{\hat I}_{\rm A} {\hat I}_{\rm B} \rangle_{\rm ent}$ for (a) number states of
Eqs. (\ref{num_1_sep}) and (\ref{num_1_ent}) with $N_1=2,N_2=0$, and (b)
coherent states of Eqs. (\ref{rho_sep_coherent}) and (\ref{rho_ent_coherent})
with $A_1=\sqrt{2},A_2=0$, against $(\omega_1-\omega_2)t$. The photon
frequencies are $\omega_1=1.2\times 10^{-4}$ and $\omega_2=10^{-4}$, in units
where $k_B=\hbar=c=1$.}
\end{figure}
\begin{figure}
\begin{center}
\scalebox{0.4}{\includegraphics{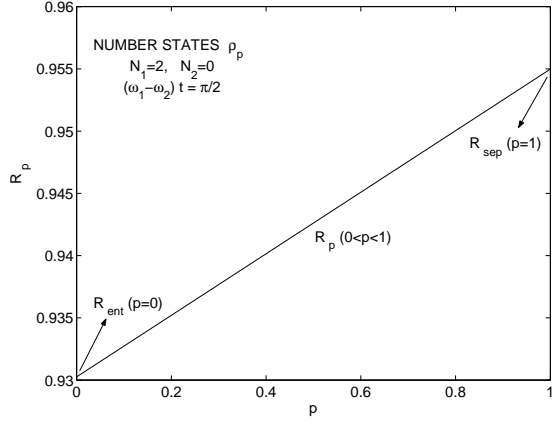}}
\end{center}
\caption{$R_{\rm p}$ against $p$ for currents that are induced by the
interpolating density matrix $\rho_{\rm p}$ of Eq. (\ref{interp}) for number
states with $N_1=2,N_2=0$. The time $t$ has been fixed so that
$(\omega_1-\omega_2)t=\pi/2$. The photon frequencies are $\omega_1=1.2\times
10^{-4}$ and $\omega_2=10^{-4}$, in units where $k_B=\hbar=c=1$.}
\end{figure}
\end{document}